\documentclass[prd,10pt,aps,superscriptaddress,nofootinbib,floatfix,twocolumn]{revtex4-1}

\usepackage{amssymb,amsmath,amsfonts,amsthm}
\usepackage{slashed, braket}
\usepackage{hyperref}

\newcommand{\tr}{\text{tr }}

\newcommand{\sign}{\text{sign}}
\newcommand{\ve}{\varepsilon}

\begin{document}

\title{Chiral Separation Effect vs. Chiral Anomaly}

\author{Z.V.Khaidukov}
\email{khaidukov.zv@phystech.edu}
\affiliation{Moscow Institute of Physics and Technology, 9, Institutskii per., Dolgoprudny, Moscow Region, 141700, Russia}
\affiliation{Institute for Theoretical and Experimental Physics of NRC ``Kurchatov Institute'', 25, B. Cheremushkinskaya, Moscow, 117259, Russia}

\author{R.A.Abramchuk}
\email{abramchuk@phystech.edu}
\affiliation{Institute for Theoretical and Experimental Physics of NRC ``Kurchatov Institute'', 25, B. Cheremushkinskaya, Moscow, 117259, Russia}

\begin{abstract}
    We study relation between Chiral Separation Effect (CSE) and Chiral Anomaly, 
    and argue that CSE does not inherit the immutability of Chiral Anomaly.
    For QED in the leading order in electric charge, with point-splitting regularization in real-time formalism, 
    we demontrate, in regularization-dependent way though,
    that CSE is an infrared (IR) effect, while the Anomaly is an ultra-violet (UV) phenomena. 
    Thus, CSE in general is sensitive to non-perturbative interactions that alter the theory IR properties, 
    while the Anomaly is fixed by Lorentz symmetry.
    The simplest example of such an interaction is the Yukawa-like fermion mass.
\end{abstract}

\date{September 16, 2022}

\maketitle

\section{Introduction}

CSE is one of the Chiral effects, which is a collective name for possible non-dissipative  currents.
The currents emerge in external fields and thermodynamic background.
The currents presumably exist in thermodynamic equilibrium, thus suggesting a new type of non-dissipative transport.

The effects are theoretically attractive because the chirality operator algebraically zeroes corrections in the closest orders in the external fields
(due to the property \(\tr \gamma_5\gamma_\mu\gamma_\nu\gamma_\lambda\gamma_\rho = 4i\epsilon_{\mu\nu\lambda\rho}\)).
Even more, the apparent relation to the corresponding anomalies, Chiral and Gravitational, seeds the idea of inherited immutability of the Chiral effects. 

Chiral anomaly is violation of a classical conservation law in QFTs.
The axial current is  conserved ``on the classical level'' (follows from the Dirac equation) for massless fermions
\begin{equation}
    j^\mu_5(x)= \bar{\psi}(x)\gamma^{\mu}\gamma^{5}\psi(x), \label{eq1}
    \quad \partial_{\mu}j^\mu_5=2im\bar{\psi}\gamma^{5}\psi.
\end{equation}  
``On the quantum level,'' however, the current operator is singular ---
it requires a regularization and further special treatment, 
which eventually leads to the renowned anomalous divergence.
The need of regularization for correct calculation of Chiral effects was noted in the original papers \cite{Vilenkin,Vilenkin2}.  

We apply the point-splitting regularization 
\begin{equation}
    j_5^{\mu}(x,\ve) = {\bar{\psi}(x+\ve/2)U(x+\ve/2,x-\ve/2)\gamma^{\mu}\gamma^{5}\psi(x-\ve/2)} \label{eq2},\\
\end{equation}
\begin{align}
    \partial_{\mu}j^{\mu 5}(x,\ve) =
    &2im\bar{\psi}(x+\ve/2)\gamma^{5}U(x+\ve/2,x-\ve/2)\psi(x-\ve/2)+ \nonumber\\
    &+i\ve^\nu j^\mu_5(x,\epsilon)F_{\nu\mu} \label{important}.
\end{align}
The parallel transporter \(U(x,y)=\exp\left(-i\int^{x}_{y}A_{\mu}dz^{\mu}\right)\) preserves the gauge invariance.
Since we do not consider renormalization in this paper, we absorb the electric charge by redefinition of the gauge field.

The chiral effects exist in certain thermodynamic background, which is indifferent for the anomaly.
In the real-time formalism the fermion propagator reads \cite{Dolan,Thermo}
\begin{gather}
    S(p)=(\slashed{p}+m)G(p),\label{prop}\\
    \quad G(p) = \frac{i}{p^{2}-m^2+i0}-2\pi\delta(p^{2}-m^2)n(|p_{0}|-\sign(p_{0})\mu)\nonumber, 
\end{gather}
where \(n(E) = (e^{E/T}+1)^{-1}\) is the Fermi-Dirac distribution.
For the present purpose, the formalism is suitable, 
since it does not discriminate the temporal direction, 
and thus allows to remove the point-splitting symmetrically.
Though QFTs in Euclidean space are usually ``better defined'' than in Minkowski space, 
the real-time consideration is more illustrative. 

In this paper we briefly investigate the relation between CSE and Chiral Anomaly 
in QED in the leading order in electric charge in real-time formalism 
applying the point-splitting regularization.
We demonstrate in regularization-dependent way that CSE and the Anomaly are not universally related.

\section{CSE and the Anomaly with point-splitting in real-time formalism}

The regularized axial current in the leading order is
\begin{align}
    j_5^{\mu}(x,\ve) = &\int\frac{d^4p}{(2\pi)^4}\frac{d^4k}{(2\pi)^4}  e^{-ikx -i(p+k/2)\ve}\times \nonumber\\
        &\times A^\lambda (k) p^\nu k^\sigma~ 4i\epsilon_{\mu\nu\lambda\sigma}~ G(p)G(p+k).
\end{align}
The central point of our derivation is the separation of the integral in independent `divergent' and `convergent' parts
that are responsible for the Anomaly and CSE, respectively
\begin{align}
    j_\mu(x,\ve) = &j_\mu^{(div)}(x,\ve) + j_\mu^{(conv)}(x,\ve), \\
    j_\mu^{(div)}(x,\ve) = &\int\frac{d^4p}{(2\pi)^4}\frac{d^4k}{(2\pi)^4} e^{-ikx -i(p+k/2)\ve}\times\nonumber\\
        &\times\frac{-2i\tilde F_{\mu\nu}(k)p^{\nu}}{(p^{2}-m^2+i0)((p+k)^{2}-m^2+i0)},\label{intdiv}
\end{align}
where 
\(\tilde F_{\mu\nu}(k) = \epsilon_{\mu\nu\lambda\sigma}F^{\lambda\sigma}(k) = -2\epsilon_{\mu\nu\sigma\lambda}ik^\sigma A^\lambda(k) \),
\begin{align}
    j_\mu^{(conv)}&(x,\ve) = \int\frac{d^4p}{(2\pi)^4}\frac{d^4k}{(2\pi)^4}  e^{-ikx -i(p+k/2)\ve}~ (-2)\tilde F_{\mu\nu}(k)p^{\nu}\times\nonumber\\
        \times\biggl(&   
        \frac{(-2\pi i)\delta((p+k)^2-m^2)n_{p+k}}{p^{2}-m^2+i0}
        + \frac{(-2\pi i)\delta(p^2-m^2)n_{p}}{(p+k)^2-m^2+i0} + \nonumber\\
        &+ (-2\pi)\delta((p+k)^2-m^2)n_{p+k}(-2\pi)\delta(p^2-m^2)n_{p}\biggr) \label{intconv}
\end{align}
Zero temperature limit of the latter equation was obtained with Kubo formula in \cite{Buiv}.

The `divergent' integral acquires most of its value in the UV limit, while the `convergent' --- in IR.
Lorentz symmetry dictates the UV properties of the theory, and thus fixes the anomaly expression.
On the other hand, the IR properties are model-specific (Dirac mass, interaction, phase transitions, etc.). 
Thus we argue for CSE dependence on interaction in the general case.
However, Chiral Anomaly can be calculated at any scale, 
and the result is the same --- the Anomaly is immutable.

The double delta term in the `convergent' part looks dangerous\footnote{
    In the paper \cite{Dolan}  authors  suggested to neglect two-delta term,  because it can be a an artifact of  the method. 
}, but equals to zero
(at least for uniform external fields, which are considered in this paper).
A way to show it is to represent delta functions as `smeared' --- large but finite at zero.

Other terms suffer from IR divergence in the external photon momentum.
The improper integrals are not absolutely convergent,
so in order to get a finite result we are 
    to be wary of the integration order and 
    to shift momentum poles to the complex plane.
The proper integration order is \(\int d^3p dp_0,\int dk_z d^2k_\bot dk_0\).
Apparently the problem is ill-defined in the infinite space\footnote{
    A thourough way to resolve the issue is to enclose the system \cite{BC} (e.g. with MIT or Bogolubov bag boundary conditions), 
    which would eliminate the troublesome long-wave modes.
}
and we are satisfied here with the simple workaround.

CSE stems from the `convergent' part solely since 
\begin{gather}
    j_\mu^{(div)}(x,\ve) ~\sim \frac{\ve_\mu}{\ve^2}\times(\text{finite value}) \xrightarrow[ \ve\to 0 ]{} 0. 
\end{gather}
The equivalence symbol `\(\sim\)' means equality up to terms that disappear in the \(\ve\to 0\) limit.
The limit is to be calculated `symmetrically' \cite{PS}.

The convergent part \eqref{intconv} at zero temperatures provides the standard CSE expression 
\begin{gather}
    j_\mu^{(conv)}(x,\ve=0) = \frac{\text{sign}~\mu\sqrt{\mu^2-m^2}}{4\pi^2}\tilde F_{0\mu} \equiv \braket{j_\mu^{5(CSE)}},\label{mCSE}
\end{gather}
where \(\braket{\ldots}\) implies averaging over the thermodynamic background in the external field.


The Anomaly follows from \eqref{important}.
The `convergent' part is unrelated since \(\ve_\nu j_\mu^{(conv)}(x,\ve) \sim 0\).

A way to calculate \(\ve_\nu j_\mu^{(div)}\) \eqref{intdiv} is to move the splitting \(\ve_\nu\) under the integral,
and substitute it with differentiation 
\(\ve_\nu \to i\frac{\partial}{\partial p^\nu}e^{-ip\ve}\).
Then we perform integration by parts to recast the derivative.
After the integration by parts, the regularization may be safely removed
\begin{align}
    \ve_\nu j^{(div)}_\mu(x,\ve) = &\int\frac{d^4p}{(2\pi)^4}\frac{d^4k}{(2\pi)^4} 
    \left(i\frac{\partial}{\partial p^\nu}e^{-ikx -i(p+k/2)\ve}\right)\times\nonumber\\ 
         &\times\frac{-2i\tilde F_{\mu\rho}(k)p^\rho}{(p^2-m^2+i0)((p+k)^2-m^2+i0)} \\
         = &\int\frac{d^4p}{(2\pi)^4}\frac{d^4k}{(2\pi)^4} e^{-ikx -i(p+k/2)\ve} \times\nonumber\\
         \times\frac{\partial}{\partial p^\nu}&\left(\frac{-2\tilde F_{\mu\rho}(k)p^\rho}{(p^2-m^2+i0)((p+k)^2-m^2+i0)}\right) \\
         \xrightarrow[ \ve\to 0 ]{} &-2\tilde F_{\mu\sigma}
         \,\int\frac{d^4p}{(2\pi)^4}\frac{\partial}{\partial p^\nu} \frac{p^\sigma}{(p^2-m^2)^2},
\end{align}
where the external field is uniform \(F_{\mu\nu}(x)=\text{const}.\)

With the regularization removed, the Gauss theorem may be applied (implying Wick rotation)
\begin{align}
    \int d^4p \frac{\partial}{\partial p^\nu}\frac{p_\sigma}{(p^2-m^2)^2} 
    &= \oint_{p\to +\infty} d^3S^\nu \frac{p_\sigma}{(p^2-m^2)^2} \nonumber\\
    &= \oint_{p\to +\infty}\frac{p_\nu p_\sigma}{p^2} d\Omega
    = g^{\nu\sigma}\frac{\pi^2}{2}.
\end{align}
\(d\Omega\) is the solid angle in 4d (\(\oint_{S^3}d\Omega = 2\pi^2\)).
Since the integral is over an arbitrary large sphere in momentum space,
only the UV properties matters --- the delta functions from thermodynamics and mass terms does not contribute,
while the linear dispersion in the UV is fixed by Lorentz symmetry. 

Finally, we obtain
\begin{gather}
    \ve^\nu j^{(div)\mu}_5(x) = -\frac{\tilde F_{\mu\nu}}{16\pi^2},\\
    \braket{\partial_\mu j^\mu_5}  
     = 2im\braket{\bar\psi\gamma_5\psi} - \frac{\alpha}{4\pi} \epsilon_{\mu\nu\lambda\sigma} F^{\mu\nu}F^{\lambda\sigma},
\end{gather}
where we reinstated the coupling constant \(\alpha=e^2/4\pi\).

\section{Simplest non-perturbative correction to CSE}

CSE for a fermion with Yukawa mass crucially depends on the coupling \(h\)
\begin{equation}
    {\cal L} =\bar{\psi}(i\slashed{D} -h\phi)\psi  
    + \frac{1}{2}\partial_\mu \phi \partial^\mu \phi - v\phi^{2} + \lambda \phi^{4}.
\end{equation}
If $ v<0,\lambda>0$ the scalar field has non-zero vacuum expectation value $\braket{\phi}=\sqrt{\frac{v}{\lambda}}$.
At \(h^*=0,\lambda^*>0\), where the scalar field is decoupled, CSE conductivity is standard 
while at \(h^*>0,\lambda^*>0\) CSE conductivity is provided with \eqref{mCSE},
where $m=h^*\sqrt{v/\lambda^*}$. 
The Anomaly is independent of the scalar field couplings.

\section{Conclusion}

With point-splitting regularization we derived the Chiral Anomaly and Chiral Separation Effect from the same starting expression.
In our derivation, the Anomaly is manifestly an UV effect, while CSE --- an IR one.
We used Yukawa model as an example of CSE receiving correction
while Anomaly is immutable.

We see the immutability of  CSE in massless case \cite{Phur} in lattice simulations as an extraordinary result,
which requires further analysis to clarify the protection mechanism on analytical level. 

In this paper we tried to show that the Chiral Effects depend on interaction and phase transitions,
since the effects at sufficiently small temperatures are saturated with states near Fermi surface.
Thus, corrections to the Chiral Effects might be investigated, for example, with Hard Loops techniques.
Also, we believe that the results of the present paper might be reproduced with Matsubara technique.

\section{Acknowledgements}

This work was supported by the Russian Science Foundation Grant Number 21-12-00237.






\bibliographystyle{plain} 
\bibliography{CSEvsAnomaly.bib}


\end{document}